\documentstyle[prl,twocolumn,aps]{revtex}

\begin{document}

\title{COHERENT CONTROL OF ISOTOPE SEPARATION IN
HD$^+$ PHOTODISSOCIATION BY STRONG FIELDS}

\author{Eric Charron$^{(1)}$, Annick Giusti-Suzor$^{(1,2)}$
and Frederick H. Mies$^{(3)}$}

\address{
$^{(1)}$ Laboratoire de Photophysique Mol\'eculaire, Universit\'e Paris-Sud,
91405 Orsay, France.\\
$^{(2)}$ Laboratoire de Chimie-Physique, 11 rue Pierre et Marie Curie, 75231
Paris, France.\\
$^{(3)}$ National Institute of Standards and Technology, Gaithersburg, Md
20899, USA.}

\address{\rm\vspace*{0.4cm}
The photodissociation of the HD$^+$ molecular ion in intense short-pulsed
linearly polarized laser fields is studied using a time-dependent wave-packet
approach where molecular rotation is fully included. We show that applying a
coherent superposition of the fundamental radiation with
its second harmonic can lead to asymmetries in the fragment angular
distributions, with significant differences between the hydrogen and deuterium
distributions in the long wavelength domain where the permanent dipole is most
efficient. This effect is used to induce an appreciable isotope separation.}

\address{\rm\flushleft PACS numbers : 33.80.-b, 33.80.Gj, 33.80.Wz}

\maketitle

Control of chemical reactivity using two-color laser fields has been actively
investigated
in the past few years.
The use of a coherent superposition of a fundamental radiation with one of its
harmonics
can induce complex interference effects between different pathways in atomic or
molecular
processes~\cite{Shapiro88}. This method is shown here to be a useful tool to
control
molecular dissociation
in the high intensity regime. Although several interesting processes have been
discovered
(above threshold dissociation \cite{Annick90,Bucksbaum}, bond softening
\cite{Bucksbaum,Jolicard92}, vibrational trapping
\cite{trapping,Bucksbaum93,Aubanel}),
they are generally in competition and are not easily disentangled in practice.
This
experimental fact motivated us to determine laser parameters which will
specifically enhance one of these physical processes. Indeed, we
have shown that in intense laser fields the branching ratios of H$_2^+$ in the
different photodissociation channels can be appreciably controlled in a
($\omega,3\omega$) experiment \cite{Eric93} and that asymmetric angular
distributions of the fragments H$^+$ and H(1s) can be created in a
($\omega,2\omega$) experiment \cite{Eric94}. Encouraged by the
efficiency of this method we decided to explore possible asymmetries in the
photodissociation of HD$^+$.

This ion presents an asymmetric electronic distribution which is small
in the bound molecule but becomes extremely important when the fragments
dissociate. This is because the two lowest electronic states, of
$^2\Sigma^+\/$ symmetry, asymptotically lead either to H$^+$+D(1s) or
H(1s)+D$^+$, with an energy difference of 29.8 cm$^{-1}$ which
is due to the larger reduced mass of the electron on the
deuterium compared to the hydrogen atom \cite{HDa,HDb}.
In the ground state, the electron is localized near the deuterium atom while
for the first excited state it is on the hydrogen. The chemical
difference between these two dissociative channels together with the inherent
simplicity of this system makes HD$^+$ one of the best candidates for
demonstrating laser-induced control of chemical reactivity.

To simulate the dissociation of the HD$^+$ molecular ion subjected to an
intense
and short laser pulse, we solve the time-dependent Schr\"odinger
equation using the short-time propagator splitting method
\cite{Fleck82}. The molecular ion interacts with a classical linearly
polarized electric field made by a coherent superposition of a fundamental
radiation of angular frequency $\omega_f\/$ and electric field amplitude
$E_f$ with its second harmonic of angular frequency $\omega_h = 2\,\omega_f\/$
and amplitude $E_h\/$
\begin{equation}
{\bf E}(t) = f(t) \left\{ E_f\,\cos(\omega_ft) + E_h\,\cos(\omega_ht+\varphi)
\right\} {\bf \hat e}\;,
\end{equation}
where $f(t) = sin^2 \left( \pi t / 2 T_p \right)$ is a Gaussian-like pulse
shape of width $T_p\/$
and total duration $2\,T_p\/$. The wavefunction describing HD$^+$ is restricted
to include only
the two first electronic states, with $\phi_G({\bf r},R)\/$ and $\phi_E({\bf
r},R)\/$ denoting
their electronic wavefunctions
\begin{equation}
\Phi({\bf R},{\bf r},t) = \left\{ F_G({\bf R},t)\,\phi_G +
F_E({\bf R},t)\,\phi_E \right\} \chi_s\;.
\end{equation}
${\bf R} = (R,\theta\/_R,\phi\/_R)\/$ is the internuclear vector joining the
hydrogen to the deuterium and {\bf r} is the electronic coordinate.
The electron spin, described by the $\chi_s\/$ function of space-fixed
projections
$s = \pm1/2\/$, acts here as a spectator and is uncoupled from the molecular
frame.
$F_G({\bf R},t)\/$ and $F_E({\bf R},t)\/$
denote the time-dependent nuclear wavefunctions associated with each electronic
state and contain the nuclear dynamic information.

To introduce the angular degrees of freedom of the molecule, we expand each
nuclear wavefunction in a large basis set of spherical harmonics
$Y_{N,M}(\theta\/_R,\phi\/_R)\/$ (typically 20 to 40 angular momenta $N\/$
are necessary to achieve convergence \cite{Aubanel}, depending on the field
intensity). We assume
HD$^+$ to be initially $(t=0)\/$ in a well defined rovibrational level
$(v_{\circ},N_{\circ},M_{\circ})\/$ of the ground electronic state :
$F_{G}({\bf R},0) = \psi_{v_{\circ},N_{\circ}}(R)
Y_{N_{\circ},M_{\circ}}(\theta\/_R,\phi\/_R)$
and $F_{E}({\bf R},0) = 0$.
For a linearly polarized electric field the initial $M_{\circ}\/$ quantum
number
is conserved, and the wavefunctions $F_G$({\bf R},t) and $F_E$({\bf R},t)
are expanded on a set of spherical harmonics with fixed azimuthal quantum
number
\begin{equation}
F_{G(E)}({\bf R},t) = \sum_N
F_{G(E),N}(R,t)\,Y_{N,M_{\circ}}(\theta\/_R,\phi\/_R)\;.
\end{equation}
In contrast to the H$_2^+$ case \cite{Eric94},
this expansion in the spherical harmonics basis includes both
parities for each electronic state, due to the {\em permanent
dipole moment\/} in HD$^+$ which allows transitions within a given
electronic state.
Indeed, the radiative interaction directly couples the nuclear wavefunctions
of components $N \leftrightarrow N \pm 1\/$. The corresponding coupling matrix
elements, evaluated in the length gauge, take the same form as for H$_2^+$
\begin{equation}
V_{N,N-1}^{\alpha\beta}(R,t) =
\mu_{\alpha\beta}(R)\,\sqrt{\frac{N^2-M_{\circ}^2}{(2N-1)(2N+1)}}\,E(t)\;,
\end{equation}
except for the expression of the different dipole moments
$\mu_{\alpha\beta}(R)\/$
between a pair of electronic states denoted by the subscripts $\alpha\/$ and
$\beta\/$ $\equiv\/$ {\small\it G\/} or {\small\it E\/}.
To take the non-degeneracy effects into account, and thus present the correct
dissociation behavior,
we use the {\em transformed\/} Hamiltonian defined by Moss
and Sadler \cite{HDa}, and employ the {\em coupled states\/} approach of
Carrington
{\em et al\/} \cite{HDb}.
We expand the resultant ground $\phi_G\/$ and first
excited $\phi_E\/$ electronic wavefunctions in terms of the pure
1s$\sigma_g\/$ and 2p$\sigma_u\/$ solutions of the Born-Oppenheimer Hamiltonian
\begin{equation}
\begin{array}{rrccccccc}
\phi_G = &  a(R) & \phi_{1s\sigma_g} + b(R) & \phi_{2p\sigma_u}\;\;
& \sim_{_{\!\!\!\!\!\!\!\!\!\!\! R\rightarrow\infty}} +\phi_{1s}({\rm D})\\
\phi_E = & -b(R) & \phi_{1s\sigma_g} + a(R) & \phi_{2p\sigma_u}\;\;
& \sim_{_{\!\!\!\!\!\!\!\!\!\!\! R\rightarrow\infty}} -\phi_{1s}({\rm H})\\
\end{array}
\end{equation}
where $a \simeq 1\/$ and $b \simeq 0\/$ for short internuclear distances.
At large distances, both $a\/$ and $b\/$ approach $1/\sqrt{2}\/$ and the
$\phi_G\/$
and $\phi_E\/$
wavefunctions correlate with the 1s atomic orbitals centered on the deuteron or
on the proton respectively.
On the way to dissociation, the electron can no more ``jump'' from one fragment
to
the other, and the electronic dipole moment $\mu_{EG}(R)\/$ vanishes at long
distance. But in this limit, the internal couplings $\mu_{GG}(R)\/$
and $\mu_{EE}(R)\/$ diverge as $-2R/3\/$ and $R/3\/$ respectively
\cite{HDa,HDb}.
The non-adiabatic couplings induced by the non-diagonal elements of the
$\partial^2/\partial R^2\/$ kinetic energy operator in the ({\small\it G,E\/})
electronic basis are also included in the time propagation using an additional
basis transformation.

At the end of the propagation, the partial nuclear wavefunctions
\mbox{$F_{G(E),N}(R,t=2\,T_p)\/$} are projected on the
corresponding field-free continuum nuclear wavefunctions to get the
dissociation
probabilities. One thus obtains the energetic and angular distributions of the
fragments \cite{Eric94}, to be compared with experimental data when available.

A simplified four-state model can be derived from the above multistate
treatment in analogy
with the two-state model commonly used for the H$_2^+$ ion in intense laser
fields \cite{trapping,Eric93} if one assumes that the molecular ion
is totally aligned by the laser electric field along its polarization axis.
This alignment has been demonstrated experimentally by different groups
\cite{align} and results from the optical pumping of the
rotational quantum number $N\/$ \cite{Eric94}.
To take the effect of the permanent dipole moments $\mu_{GG}(R)\/$ and
$\mu_{EE}(R)\/$ into account, we associate {\em two\/} nuclear wave-packets
with each electronic potential, representing the collective sum of the {\em
odd\/}
({\em o\/}) and the {\em even\/} ({\em e\/}) partial wave components in Eq.(3).
The
total wavefunction in Eq.(2) can then be expressed simply as follows
\begin{equation}
\Phi = \left\{ \left( F_{G,o} + F_{G,e} \right) \phi_G
 + \left( F_{E,o} + F_{E,e} \right) \phi_E \right\} \chi_s\;.
\end{equation}
We propagate an initial $(t = 0)\/$ wavepacket taken as a pure
$v_{\circ}\/$ vibrational state of the ground electronic potential.
The $\mu_{GG}(R)$, $\mu_{EE}(R)$ and $\mu_{EG}(R)$ dipole moments
are included in the potential matrix as
off-diagonal terms that couple nuclear wavefunctions of opposite parity.
This simple model saves a lot of time and memory in computation
and has been checked to give reliable results.
The angular distribution of the H$^+$ fragment can be obtained from the
projections
$C_{G,o}(\varepsilon)\/$ and $C_{G,e}(\varepsilon)\/$ of the odd and even
{\small\it G\/} nuclear wavepackets at the end of the pulse on the {\small\it
G\/} field-free
continuum nuclear wavefunction at the energy $\varepsilon\/$ by
\begin{equation}
\begin{array}{rcll}
P_{\rm H^+}(\varepsilon,\theta=0^\circ)   & = &
\frac{1}{2}\left|C_{G,o}(\varepsilon)-C_{G,e}(\varepsilon)\right|^2 &\\
P_{\rm H^+}(\varepsilon,\theta=180^\circ) & = &
\frac{1}{2}\left|C_{G,o}(\varepsilon)+C_{G,e}(\varepsilon)\right|^2 & ,
\end{array}
\end{equation}
where $\theta = 0^\circ$ and $\theta = 180^\circ$ denote the {\em forward\/}
and {\em backward\/}
ion signals \cite{Eric93,Eric94}. The directional dissociation probabilities
are
obtained by energy integration
\begin{equation}
P_{\rm H^+}(\theta=0^\circ\,or\,180^\circ) = \int P_{\rm
H^+}(\varepsilon,\theta=0^\circ\,or\,180^\circ)\,d\varepsilon\;,
\end{equation}
and the total dissociation probability $P_{\rm H^+}\/$ in the ground electronic
state is just the sum of these two directional probabilities.
Similar equations (with opposite signs in Eq.(7)) relative to the excited part
{\small\it E\/} of the wavepacket give
the partial and total productions of D$^+$ ions.

The most remarkable effect on the angular distributions of the HD$^+$
photofragments
is obtained using a coherent superposition of the second (or any even)
harmonic with a fundamental laser frequency. In this case a given energy of
dissociation
$\varepsilon\/$ can be reached by absorbing {\it odd\/} as well as {\it even\/}
number
of photons. Thus, for example, $C_{G,o}(\varepsilon)\/$ and
$C_{G,e}(\varepsilon)\/$ can
simultaneously contribute to the probabilities in Eq.(7), and the cross-term
Re$\left\{C_{G,o}^{*}(\varepsilon)\,C_{G,e}(\varepsilon)\right\}\/$
induces different dissociation probabilities in the $\theta=0^{\circ}\/$ and
$\theta=180^{\circ}\/$ directions. This effect is similar to the asymmetric
angular
distributions of photoelectrons observed in two-color high intensity
atomic photoionization \cite{atom}, or to the asymmetric proton angular
distributions in H$_2^+$ photodissociation \cite{Eric94}. However,
what is unique here
is that we can also induce different asymmetries in the distributions of
the H$^+$ and D$^+$ photofragments and control the spatial separation
of these isotopes. This effect requires long wavelength
lasers, such as the CO$_2$ laser case shown in figure 1 using
$\lambda_f=10.6\,\mu$m ($\hbar\omega_f=970\,$cm$^{-1}$) and
$\lambda_h=5.3\,\mu$m ($\hbar\omega_h=1940\,$cm$^{-1}$).

This figure presents the fully converged 3D results obtained from an expansion
using up to 40 coupled partial waves in each electronic potential (see Eq.(3)),
for an initial \mbox{$(v_{\circ}=0,\,N_{\circ}=0,\,M_{\circ}=0)\/$}
rovibrational state and
using peak intensities $I_f=5 \times 10^{13}\,$W.cm$^{-2}$ and
$I_h=2 \times 10^{13}\,$W.cm$^{-2}$ to evaluate the electric fields in
Eq.(1). The pulse duration is 225 fs, corresponding
to 13 optical cycles of the second harmonic. For each chosen phase $\varphi\/$
the {\em forward\/}
and {\em backward\/} dissociation probabilities for a particular fragment
are obtained by angular integration of the 3D distributions over
\mbox{$0^{\circ} \leq \theta \leq 90^{\circ}$} and
\mbox{$90^{\circ} \leq \theta \leq 180^{\circ}$} respectively, where $\theta\/$
denotes the angle between the dissociation and the polarization axis.

\vspace*{3cm}

{\small
\noindent
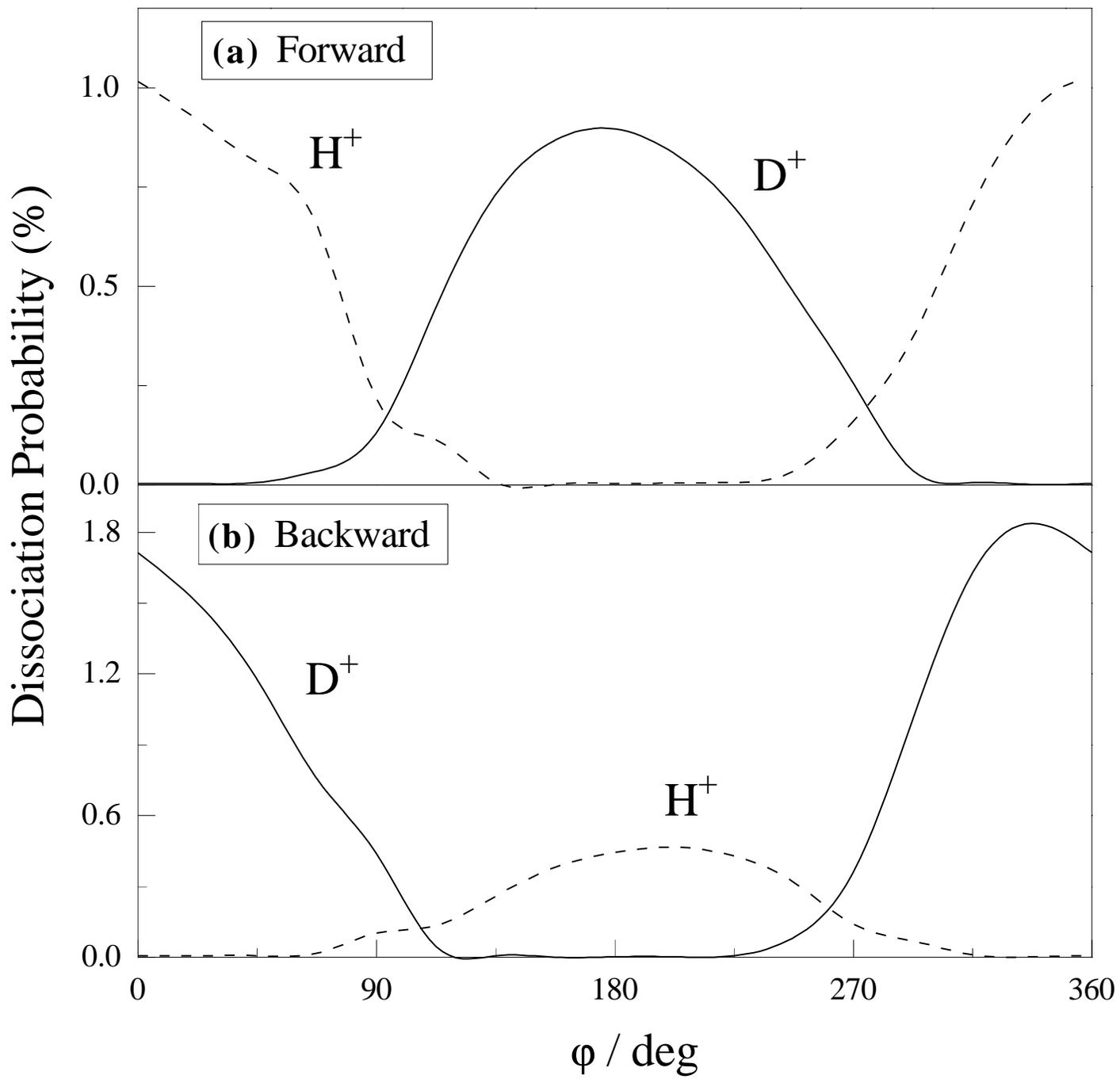
FIG. 1.
Dissociation probabilities of HD$^+$ in the two first electronic states leading
to the \mbox{H$^+$ + D(1s)} (dashed line) and \mbox{H(1s) + D$^+$}
(solid line) fragments as a function of the phase shift $\varphi\/$
between the two laser harmonics. The integrated forward
($\theta \approx 0^\circ\/$)
probabilities are shown in {\bf (a)}, and the integrated
backward ($\theta \approx 180^\circ\/$) probabilities are shown in
{\bf (b)}, for a laser
pulse of \mbox{225 fs} duration made by the coherent superposition of the two
wavelengths $\lambda_f=10.6\,\mu$m and $\lambda_h=5.3\,\mu$m with the
intensities $I_f = 5 \times 10^{13}\,$W.cm$^{-2}\,$ and
$I_h = 2 \times 10^{13}\,$W.cm$^{-2}\,$. The HD$^+$ molecular ion
is initially in the
($v_{\circ}=0,\,N_{\circ}=0,\,M_{\circ}=0$) rovibrational state.
}

\vspace*{0.3cm}
The polar plot in figure 2 for the specific phase
$\varphi=0\/$ confirms that the angular distributions are closely
peaked along the polarization axis $\hat{e}\/$.
For this phase shift $\varphi=0\/$, essentially 100\% of the
dissociated ions emitted in the {\em forward\/} direction
are expected to be H$^+$, and the D$^+$ ions are almost only ejected
in the {\em backward\/} direction.
Obviously, this gives a very efficient mechanism for spatially
separating the isotopes.

In this dissociation scheme, a neutral atom H(1s) or D(1s) always
appears in the opposite direction to the other ion (D$^+$ or H$^+$).
A significant ionization of the neutral atoms on the way to the detector should
thus just enhance the ion signal, and increase the efficiency of the separation
process.
If the phase shift is changed to $\varphi=\pi\/$
the direction of ejection of the H$^+$ and D$^+$ ions are simply reversed,
due to a change in the sign of the cross-term in Eq.(7). It is just near
$\varphi=\pi/2\/$ that the isotopes are not separated.

The explorations we have made show that this separation effect is
robust enough to survive experimental uncertainties in the intensities
of both harmonics and in their phases. Increasing the pulse length from
225 fs to 675 fs has a negligible effect on the angular asymmetry for
$\varphi=0\/$. We further tested the experimental feasibility by
performing the same calculations for a wide range of initial vibrational
and rotational states and found the same qualitative behavior as in
figure 1 for $v_{\circ}=0,1,2,3\/$ and $N_{\circ}=0,1,2$ and $M_{\circ}=0,\pm
1$.
This coherent control should thus persist even after averaging over
an initial distribution of the HD$^+$ states, as expected in a real experiment.

\vspace*{3cm}

{\small
\noindent
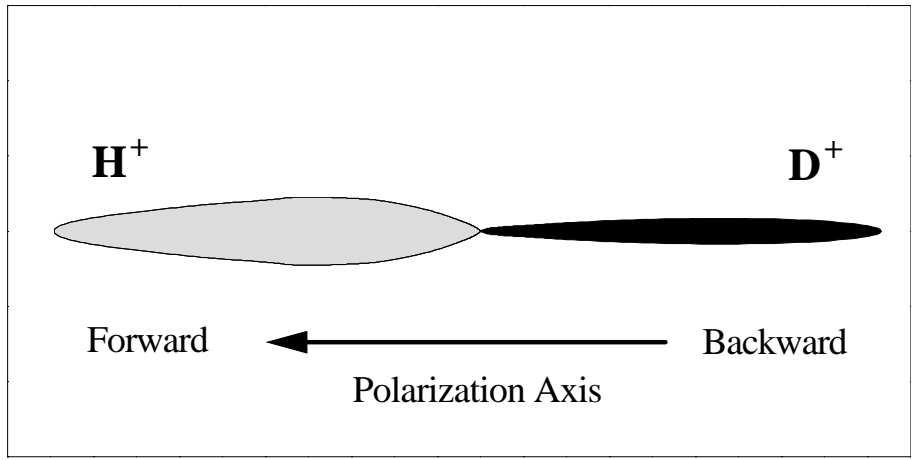
FIG. 2.
Polar plot showing the angular distributions of the H$^+$ and D$^+$
ions with the same conditions as in figure 1 for the specific phase
$\varphi=0\/$. Note the better alignment of D$^+$, due to a
more efficient rotational pumping \cite{Eric94} in the excited
electronic state.}

\vspace*{0.3cm}
The most sensitive parameter for achieving coherent control of the isotope
separation
in HD$^+$ photodissociation is the laser frequency $\omega_f\/$ in Eq.(1).
Indeed, twice the CO$_2$ laser photon energy (1940$\,$cm$^{-1}$) is comparable
to the
vibrational
spacing in HD$^+$ ground electronic state. A choice of laser frequency in this
range greatly enhances the efficiency of the $\mu_{GG}(R)\/$ and
$\mu_{EE}(R)\/$ permanent transition dipoles.
The variations of the {\em forward\/} and {\em backward\/}
dissociation probabilities as a function of the fundamental wavelength used
in this two-color study are shown in figure 3 for the most favorable phase
$\varphi=0\/$. These 1D calculations are presented for the initial level
$v_{\circ}=3\/$ using the same intensities as in \mbox{figure 1}. The
remarkable
separability seen in figure 2 persists over the full range of frequencies
\mbox{$500\,{\rm cm}^{-1} \leq \hbar\omega_f \leq 2500\,{\rm cm}^{-1}\/$.}

It is interesting to note that there are two distinct regions of enhanced
probability of photodissociation in figure 3 which we interpret as low
and high frequency mechanisms. The first one is analogous to the static field
{\em tunneling regime\/} found in strong field photoionization and
photodissociation
\cite{Corkum}, while the second is closer to the Floquet {\em multiphoton
regime\/}. The long wavelength (low frequency)
mechanism, where appreciable dissociation occurs during a single optical
cycle \cite{Corkum}, favors the spatial separation of the isotopes.
In this regime, pathways induced by the permanent dipoles $\mu_{GG}\/$
and $\mu_{EE}\/$, which have {\em opposite\/} signs, manifest interferences
with those due to the $\mu_{EG}\/$ transition dipole, and it
allows us to achieve this remarkable control.
On the contrary, the fragmentation dynamics in the short wavelength
(high frequency) domain is characterized by two features which
destroy this interference effect. Firstly, because nothing significant
happens during one optical period, the dissociation dynamics reflects
behavior ``averaged'' over the optical cycle. Secondly,
unfavorable Franck-Condon factors between nuclear wavefunctions
differing by the energy of one photon within a given electronic
state reduce the effect of permanent dipoles. In fact,
since the $\mu_{EG}\/$ coupling
dominates the dissociation in this high frequency limit,
the dynamics of HD$^+$ is very similar to the one of
H$_2^+$ and
no appreciable difference
between the H$^+$ and D$^+$ partial probabilities can be induced
in this regime,
as observed by Sheehy, Walker and \mbox{DiMauro \cite{Sheehy}}.

\vspace*{3cm}

{\small
\noindent
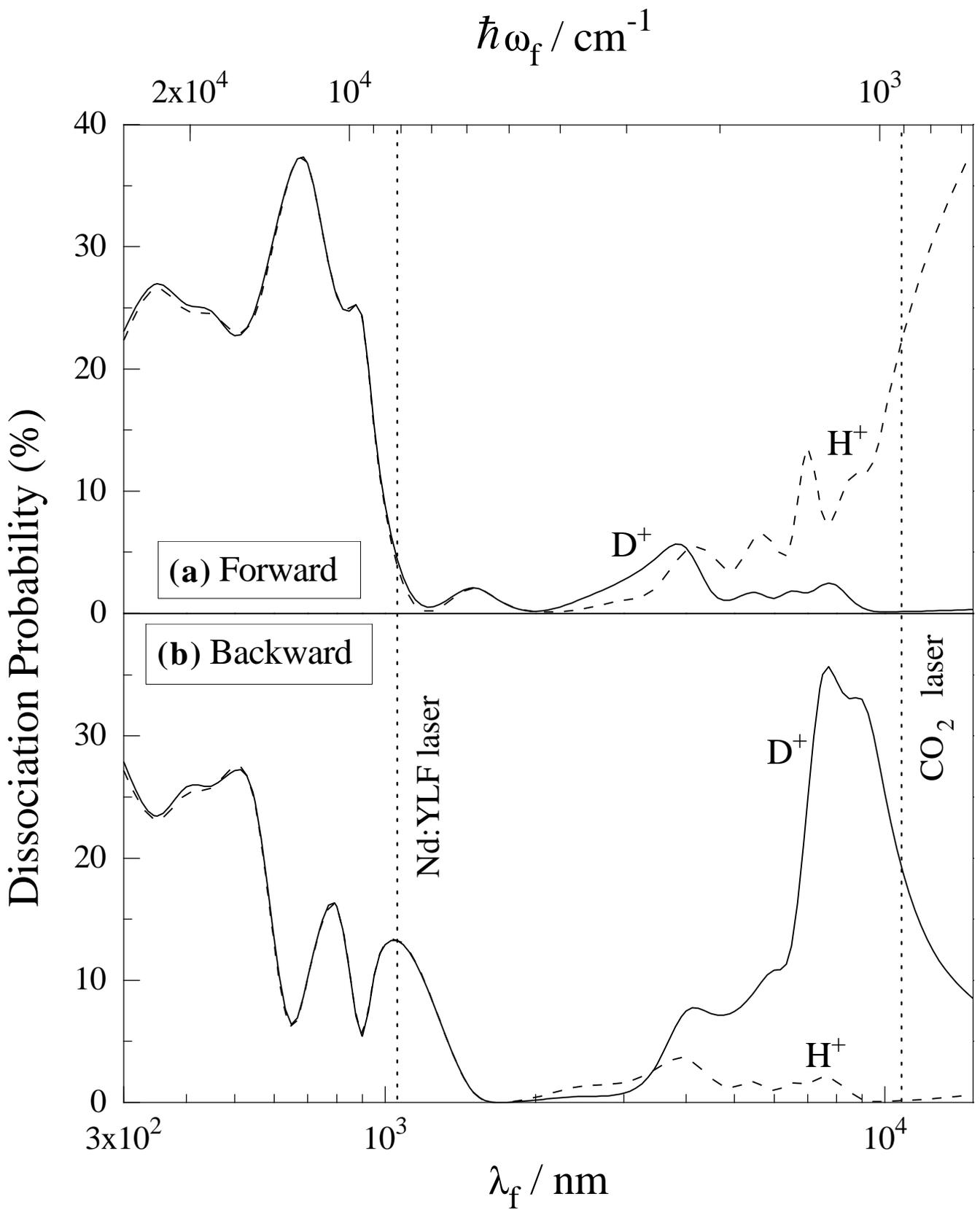
FIG. 3.
1D-dissociation probabilities of HD$^+$ ($v_{\circ}=3$) in the forward
{\bf (a)} and backward {\bf (b)} directions leading
to the \mbox{H$^+$ + D(1s)} (dashed line) and \mbox{H(1s) + D$^+$}
(solid line) fragments as a function of the fundamental wavelength.
The phase of the second harmonic is fixed to $\varphi=0$.
The pulse duration and laser intensities are the same as in figure 1.
}

\vspace*{0.3cm}
In this paper we have predicted a new and efficient mechanism for
separating different isotopic products in a fragmentation
process.
Unlike previous examples of coherent control which only
controlled the distribution of all the charged particles
together, we are now able to distinguish the fragments of
different masses.
This scenario should extend far beyond the specific example
of HD$^+$ and can be applied to separating products of a different chemical
nature as well. Tailoring quantum interferences induced
by two-color coherent light opens a promising avenue for controlling
the dynamics of quasi-symmetric systems, and applications of primary
importance can be expected in enantio-differentiation for
instance~\cite{ScAm}.

We thank B. Sheehy, B. Walker and L. DiMauro for sending us experimental
results
prior to publication. These results were of invaluable assistance in assessing
the validity of our numerical simulations.
This work was supported in part by a NATO grant for International Collaborative
Research. The computer facilities have been provided by IDRIS of the CNRS.

\end{document}